\documentclass[graybox]{svmult}

\usepackage{mathptmx}       
\usepackage{helvet}         
\usepackage{courier}        
\usepackage{type1cm}        
\usepackage{makeidx}         
\usepackage{graphicx}        
\usepackage{multicol}        
\usepackage[bottom]{footmisc}

\makeindex             

\begin{document}

\title*{Study of photon detection efficiency and position resolution of BESIII electromagnetic calorimeter}
\author{Vindhyawasini ~Prasad$^1$, Chunxiu ~Liu$^1$, Xiaobin ~Ji$^1$, Weidong ~Li$^1$, Huaimin ~Liu$^1$ and Xinchou ~Lou$^1$}

\institute{ $^1$ Experimental Physics Division, Institute of High Energy Physics, Beijing, 100049, China, \at
\email{vindy@ihep.ac.cn (Vindhyawasini ~Prasad)}}

%
%
\maketitle

\abstract{We study the photon detection efficiency and position resolution of the electromagnetic calorimeter (EMC) of the BESIII experiment. The control sample of the initial-state-radiation (ISR) process of $e^+e^-\rightarrow \gamma \mu^+\mu^-$ is used at $J/\psi$ and $\psi(3770)$ resonances for the EMC calibration and photon detection efficiency study. Photon detection efficiency is defined as the predicted photon, obtained by performing a kinematic fit with two muon tracks, matched with real photons in the EMC.  The spatial resolution of the EMC is defined as the separation in polar ($\theta$) and azimuthal ($\phi$) angles  between charged track and associated cluster centroid on the front face of the EMC crystals. }

\section{Introduction}
\label{sec:1}
The BESIII experiment is an electron-positron collider experiment running at tau-charm region located at the Institute of High Energy Physics, Beijing, China \cite{bes3_nim}. It has collected the data at the center-of-mass energies between ($2.0 - 4.6$) GeV,  more than 130 energy points, to study the Hadron spectroscopy of charm mesons and search for new physics phenomena. The BESIII uses a CsI(Tl) based electromagnetic calorimeter (EMC) to measure the electromagnetic showers of electrons and photons with excellent efficiency, energy and angular resolutions. The capability of EMC also allows to detect and measure the photons, which are used to reconstruct the $\pi^0$ and $\eta^0$ mesons as well as the decay processes where the photons are produced directly via radiative process, such as $J/\psi \rightarrow \gamma \eta \pi\pi$.

The BESIII spectrometer is a general purpose detector, covering the $93\%$ of the total solid angle, described in \cite{bes3_nim}. It's EMC is made of 6240 CsI(Tl) crystals and  has one barrel and two end-cap regions. The length of each crystal is 28 cm or $15.1 X_{0}$, where $X_{0}$ is the radiation length. The barrel region contains 5,280 crystals which are divided into 44 rings, denoted by $\theta_{index}$, and thus each ring contains 120 crystals. All the crystals are tilted within $1.5$ degree in $\phi$ directions and $1.5 - 3.0$ degrees in the $\theta$ direction ($\pm 5$ cm away from the interaction point in the beam direction) to avoid photons from the interaction point escaping through cracks between crystals. Each end-cap region contains 6 rings. The number of crystals in the six rings of each end-cap is 96, 96, 80, 80, 64 and 64. The end-caps contain 33 different sizes of crystals, and among the 960 crystals in the end-cap regions, 192 crystals are irregular pentagons. 

\section{Photon cluster shapes calibration and photon detection efficiency study}
\label{sec:2}
\begin{figure}
\sidecaption
\includegraphics[scale=.36]{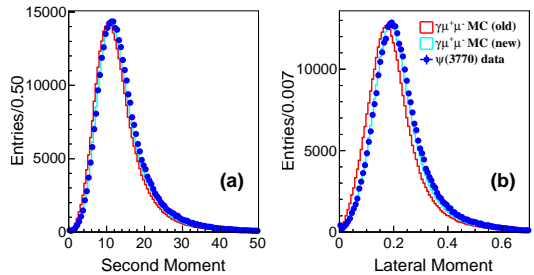}
\caption{The plots of (a) second moment and (b) lateral moment distributions for data (blue), old MC  (red) and new MC (cyan).  The old (new) MC includes the EINC value of 0.20 MeV (0.27 MeV). The distributions of second and lateral moments in the new MC describe well with the data.}
\label{fig:EINC}       
\end{figure}

We use the radiative muon pair events from the initial state radiation process of $e^+e^- \rightarrow \gamma \mu^+\mu^-$ at $J/\psi$ and $\psi(3770)$ resonances to study the photon reconstruction in EMC as well as to measure the energy and angular resolutions. The BESIII has collected  $2.9$  ($0.4$) fb$^{-1}$ of data at $\psi(3770)$ ($J/\psi$) resonance during 2009-2012. The energy and angle of the radiated photon are predicted using the 4-momenta of two charged tracks, without using any calorimeter information. The clean sample of raditive muon pair events is  also used to calibrate the cluster shapes of the photon, such as the lateral moment \cite{latmom}  and second moment, which are used to differentiate between the real and fake photons.  The lateral moment \cite{latmom} quantifies the transverse shower shape of the cluster. The second moment of shower energy is defined as $\sum_{i} E_i r_i^2$, where $E_i$ is deposited energy in the $i^{th}$ crystal and  $r_i$ the radial distance of crystal $i$ from the cluster center. The cluster shapes of these variables are adjusted while tuning the value of EMC incoherent noise (EINC) \cite{EINC} (Figure~\ref{fig:EINC}). The data in these variables seem to describe well with MC for the EINC value of 0.27 MeV. Previously, BESIII has used an EINC value of 0.20 MeV in the MC simulation.

\begin{figure}
\sidecaption
\includegraphics[scale=.36]{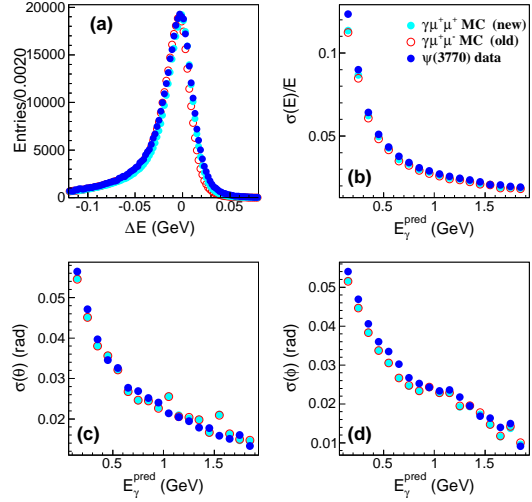}
\caption{The plots of (a) $\Delta E$ distribution, (b) energy resolution ($\sigma_{E}/E$) vs. $E_{\gamma}^{pred}$, (c) polar angle resolution ($\sigma_{\theta}$) vs. $E_{\gamma}^{pred}$ and (d) azimuthal angle resolution ($\sigma_{\phi}$) vs. $E_{\gamma}^{pred}$ for  new MC (cyan), old MC (red) and data (blue). The  $\Delta E$  is defined as the energy difference between predicted and reconstructed photons. The photon is predicted with a kinematic fit using the 4-momenta of two charged tracks, without using any calorimeter information.   }
\label{fig:EINCpsip}       
\end{figure}

We study the effect of EINC in the energy  and angular resolutions, defined as difference in energy and angular distributions between predicted and reconstructed photons, at different values of predicted photon energy ($E_{\gamma}^{pred}$) using both old and new MCs simulated with the EINC value of 0.20 MeV and 0.27 MeV, respectively (Figure~\ref{fig:EINCpsip}). The energy resolution of photon in the new MC seems to  be little bit worse than the older MC, but it is much closer to the real data. The effect of increased EINC value in the angular distributions of the EMC seems to be negligible.

\begin{figure}[!htb]
\centering
\includegraphics[scale=.62]{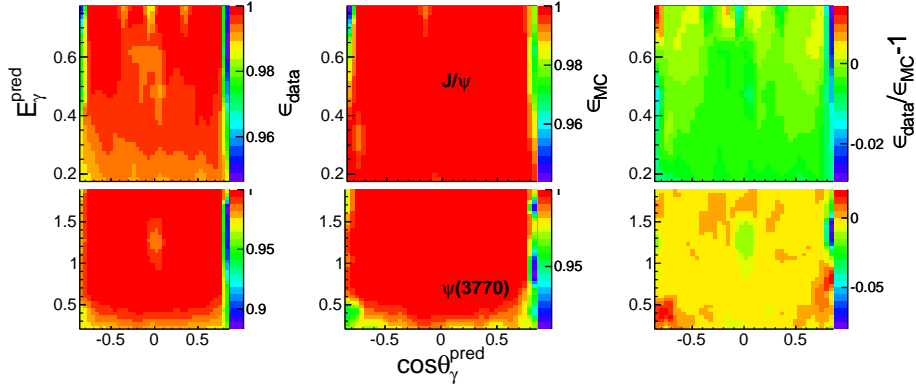}

\caption{The 2-d plots of the efficiency curves in $Cos\theta_{\gamma}^{pred}$ and  $E_{\gamma}^{pred}$ planes for data (left) and MC (middle), and the relative difference in efficiency between data and MC in $Cos\theta_{\gamma}^{pred}$ and  $E_{\gamma}^{pred}$ planes (right), where $\theta_{\gamma^{pred}}$ and $E_{\gamma}^{pred}$ are the polar angle and energy, respectively of the predicted photon obtained from the kinematic fit. Top plots are for $J/\psi$ data and bottom plots for $\psi(3770)$ data.}
\label{fig:2djpseff}
\end{figure}

We finally compute the photon detection efficiency, which is defined as the fraction of predicted photon matched with actual photons in the EMC. The $\theta_{ \gamma,\gamma^{pred}}$ and $E_{\gamma}/E_{\gamma}^{pred}$ distributions are expected to peak at 0 radian and 1, respectively, for an efficient photon, where $\theta_{ \gamma,\gamma^{pred}}$ is the angle between predicted and reconstructed photons,  and $E_{\gamma}/E_{\gamma}^{pred}$ the energy ratio of the reconstructed photon to the predicted photon. The systematic uncertainty due to photon reconstruction, defined as the relative difference in photon detection efficiency between data and MC, is found to be up to the level of $1\%$ (Figure~\ref{fig:2djpseff}).

\section{Position resolution}
The spatial resolution of the BESIII EMC is affected by the systematic shift of the reconstructed coordinate due to using a center-of-gravity technique \cite{Brabson, He} to measure the shower position of the particles in the EMC. The shift value depends on the position of the entry point relative to the center of the nearest crystal. We can correct the shift effect in two independent coordinates of polar ($\theta$) and azimuthal ($\phi$) angles, which are shifted after the reconstruction and thus degrade the position resolution of the calorimeter. We use single electron (photon) events for polar (azimuthal) angle correction in the front face of the EMC where the incident particle is expected to create a shower. The correction parameter is defined as the difference between the cluster centroid and expected positions of the track in the front face of the EMC. The correction is performed in both barrel and end-cap regions in the three different energy regions. In the barrel region, we also perform the correction factor study for the twenty two values of $\theta_{index}$ while assuming the first 22 rings to be the mirror image of the rest of the other 22 rings. 

\begin{figure}
\sidecaption
\includegraphics[scale=.31]{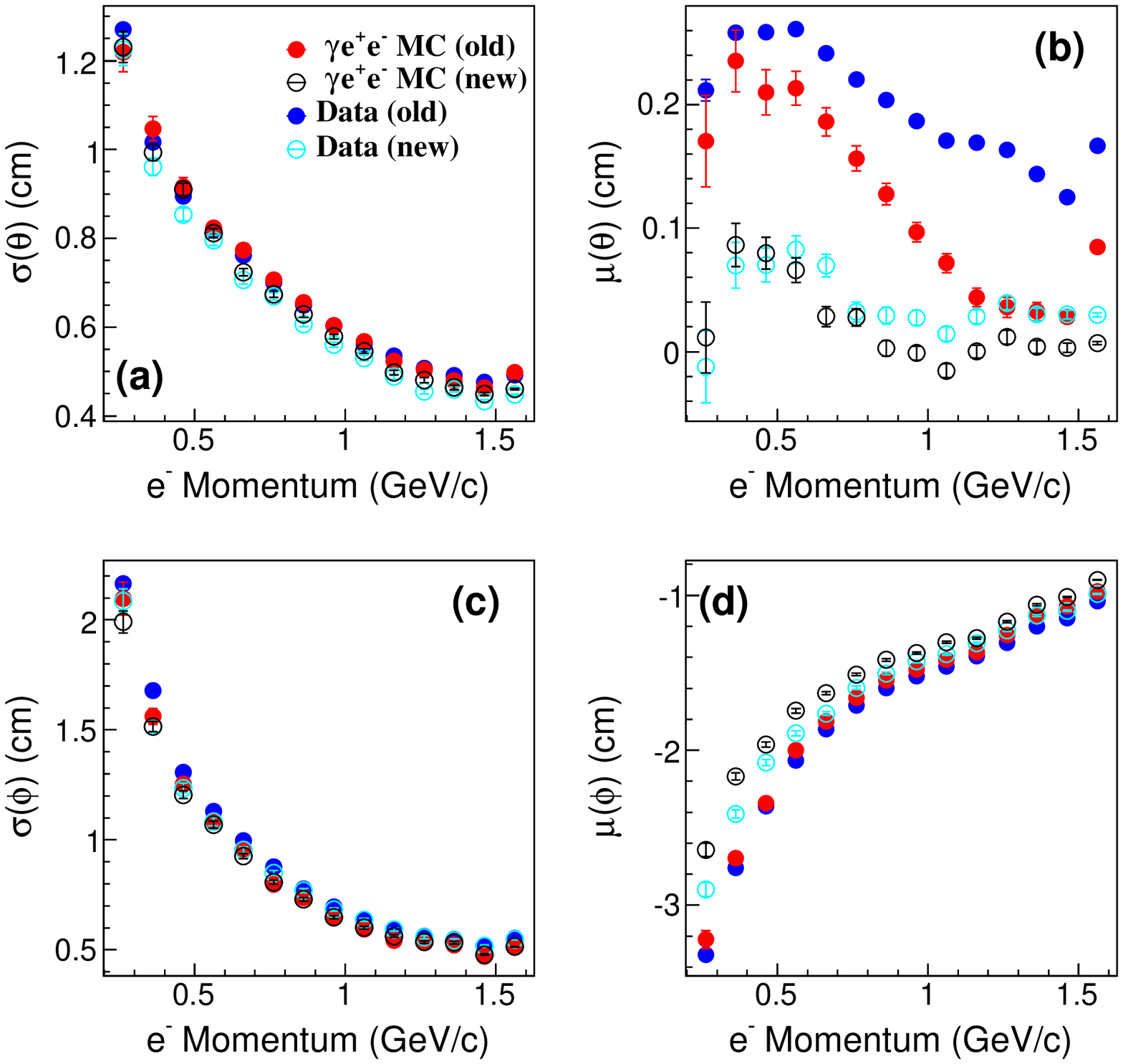}
\caption{The resolution (a,c) and mean (b,d) of $\Delta \theta$ and $\Delta \phi$  distributions as a function of $e^-$ momentum.  Top plots are for $\Delta \theta$ distribution and bottom plots  for $\Delta \phi$  distribution. The new calibration parameters seem to improve the position resolution of EMC and minimize the bias in the measurement of the shower position of the particles in the EMC. The resolution of $\Delta \phi$ distribution in the low-momentum region seems to be worse due to the effect of magnetic field. }
\label{performance}       
\end{figure}

The performance of new correction parameters is checked using a radiative Bhabha sample at $J/\psi$ resonance in  $\delta \theta$ and $\delta \phi$ distributions, defined as separations in ($\theta,\phi$) positions between cluster centroid and the expected position of the track on the front face of the EMC. The angular distributions are further translated as $\Delta \theta \rightarrow \delta \theta \times l$ and $\Delta \phi \rightarrow \delta \phi \times r$, where $l$ is the  path length of extrapolated track between interaction point and EMC cluster centroid, and $r$ is the inner radius of the EMC. Now the units of $\Delta\theta$ and $\Delta \phi$ are measured in cm.  Figure~\ref{performance} shows the resolution and mean values of $\Delta \theta$ and $\Delta \phi$ distributions as a function of $e^-$ momentum. The resolution of $\Delta \phi$ distribution seems to be a little bit worse in the low-momentum region due to the effect of magnetic field. The new correction parameters seem to improve the mean and sigma values of $\Delta \theta$ and $\Delta \phi$ distributions.

We also use a control sample of $J/\psi \rightarrow \rho\pi$ decay to check further the performance of new EMC position correction parameters. The di-photon invariant mass in this decay process is defined as, $m_{\gamma \gamma} = \sqrt{2E_{low}E_{high}(1-cos\theta_{\gamma\gamma})}$, where $E_{low}$ and $E_{high}$ are the low and high energies of the photons from $\pi^0$ decay and $\theta_{\gamma\gamma}$ is the angle between them. Table~\ref{tab:mgg} summarizes the mean and sigma values, calculated using an asymmetric Gaussian function, of $m_{\gamma\gamma}$ distribution for data and MC produced while using new and old EMC correction parameters. A marginal improvement in the resolution of $m_{\gamma \gamma}$ distribution seems to be achieved using the new correction parameters.

\begin{table}
\caption{Mean and $\sigma$ values of $m_{\gamma\gamma}$ distribution in $J/\psi \rightarrow \rho\pi$ decay for old and new data and MCs.}
\label{tab:1}       
%
%
\begin{tabular}{p{2cm}p{2.4cm}p{2cm}p{2.4cm}p{2.4cm}}
\hline\noalign{\smallskip}
Parameters   & old MC & new MC  & old data & new data  \\
\noalign{\smallskip}\svhline\noalign{\smallskip}
mean (MeV/$c^2$) & $134.33 \pm 0.04$   & $134.32 \pm 0.04$ & $134.52 \pm 0.02$ &$134.49 \pm 0.02$ \\
$\sigma$ (MeV/$c^2$) & $4.30 \pm 0.04$  & $4.28 \pm 0.05$ & $4.25 \pm 0.03$ & $4.20 \pm 0.02$\\

\noalign{\smallskip}\hline\noalign{\smallskip}
\end{tabular}
\label{tab:mgg}
\end{table}

\section{Summary and conclusion}
We study the photon detection efficiency and position resolution of the BESIII electromagnetic calorimeter. The relative differences in photon detection efficiencies between data and MC are observed up to the level of $1\%$. The position resolution of the EMC has been studied in two independent coordinates of polar and azimuthal angles. The new correction parameters improve the spatial resolution of the EMC. A marginal improvement in $m_{\gamma\gamma}$ distribution for $J/\psi \rightarrow \rho\pi$ decay process is also achieved while using the new correction parameters. The new EMC correction parameters have been included in the new releases of the BESIII software packages.

\end{document}